\documentclass[onecolumn, doublespace, 12pt fonts]{aastex6}

\usepackage{graphicx,latexsym,amssymb, epsfig}
\usepackage{multirow,amsmath,array,booktabs}
\usepackage{subfigure}
\usepackage{tabularx}
\usepackage{color}
\usepackage{bm}
\usepackage[figuresright]{rotating}
\usepackage[section]{placeins}
\usepackage{slashed}
\usepackage{graphicx}
\usepackage{tikz}

\newcommand{\nc}{\newcommand}       % new command
\nc{\vc}[1] {\mbox{\boldmath $#1$}} % boldmath(vector)
\nc{\del}       {\partial}              % bra state
\nc{\bra}       {\langle}               % bra state
\nc{\ket}       {\rangle}               % ket state
\nc{\bras}[1]   {\langle #1|}           % bra state
\nc{\kets}[1]   {|#1\rangle}            % ket state
\nc{\mapleft}[1]{           % something under arrow
	\smash{\mathop{\,          %
			\hbox to 1.5cm{\rightarrowfill}\, }\limits_{#1}}}
\nc{\beq}     {\begin{eqnarray}} \nc{\eeq}    {\end{eqnarray}}
\nc{\nn}      {\\\nonumber} \nc{\vs}      {\vspace{-0.275cm}}
\nc{\fra}    {\frac{1}{2}}
\nc{\mb}        {\mathbf}

\begin{document}
	\title{The possibility of the secondary object in GW190814 as a neutron star}
	\author{Kaixuan Huang\altaffilmark{1}, Jinniu Hu,\altaffilmark{1,2}, Ying Zhang\altaffilmark{3,2}, Hong Shen\altaffilmark{1}}
	\altaffiltext{1}{School of Physics, Nankai University, Tianjin 300071,  China}
	\altaffiltext{2}{Strangeness Nuclear Physics Laboratory, RIKEN Nishina Center, Wako, 351-0198, Japan}
	\altaffiltext{3}{Department of Physics, Faculty of Science, Tianjin University, Tianjin 300072, China}

	\email{hujinniu@nankai.edu.cn; yzhang@tju.edu.cn;songtc@nankai.edu.cn}

	\begin{abstract}
    A compact object was observed with a mass $2.50-2.67~M_\odot$ by LIGO Scientific and Virgo collaborations (LVC) in GW190814, which provides a great challenge to the investigations into the supranuclear matter. To study this object, properties of neutron star are systematically calculated within the latest density-dependent relativistic mean-field (DDRMF) parameterizations, which are determined by the ground state properties of spherical nuclei. The maximum masses of neutron star calculated by DD-MEX and DD-LZ1 sets can be around $2.55~M_\odot$ with quite stiff equations of state generated by their strong repulsive contributions from vector potentials at high densities. Their maximum speeds of sound $c_s/c$ are smaller than $\sqrt{0.8}$ at the center of neutron star  and the dimensionless tidal deformabilities at $1.4~M_\odot$ are less than $800$. Furthermore, the radii of $1.4 ~M_\odot$  also satisfy the constraint from the observation of mass-radius simultaneous measurements (NICER). Therefore, we conclude that one cannot exclude the possibility of the secondary object in GW190814 as a neutron star composed of hadron matter from DDRMF models. 
	\end{abstract}
	\keywords {GW190814 - Neutron Star - DDRMF model - Gravitational Waves}
	
	\section{Introduction}
		% A review on the neutron star 
		The rapid progresses of the astronomical observable techniques provide not only great challenges but also many opportunities in the investigations of neutron star. In the past decade, the measurements of massive neutron stars successively  broke through our recognition of their maximum masses from PSR J1614-2230 ($1.928\pm0.017~M_\odot$)~\citep{demorest2010,fonseca2016}, PSR J0348+0432 ($2.01\pm0.04~M_\odot$)~\citep{antoniadis2013}, to PSR J0740+6620 ($2.14_{-0.09}^{+0.10}~M_\odot$)\\~\citep{cromartie2020}. For the first time, the mass and radius of PSR J0030+0451 were simultaneously measured by the Neutron star Interior Composition Explorer (NICER) collaboration who drew the first-ever map of neutron star~\citep{raaijimakers2019}. Its was reported to have  a mass of $1.44_{-0.14}^{+0.15}~M_\odot$ with a radius of $13.02_{-1.06}^{+1.24}$ km ~\citep{miller2019} and a mass of $1.34_{-0.16}^{+0.15}~M_\odot$ with a radius of $12.71_{-1.19}^{+1.14}$ km~\citep{riley2019} by two independent analysis groups.
		
		At the same time, the multi-messenger astronomy era has begun with the successful operation of gravitational wave detectors, LIGO Scientific and Virgo Collaborations (LVC), which firstly received the gravitational waves generated by a binary neutron-star (BNS) merger GW170817 event~\citep{abbott2017a, abbott2017b, abbott2018}. The tidal deformability of neutron star was estimated from the signals, which becomes a new constraint on the equation of state of neutron star matter. The total mass of this BNS system in GW170817 is around $2.7~M_\odot$ and the mass of the heavier component is around $1.16-1.60 ~M_\odot$ with lower-spin priors, while the maximum mass of neutron star can approach $1.89 ~M_\odot$ with high-spin priors~\citep{abbott2019}.  After that, the second possible BNS merger was observed in April of 2019 as GW190425, with the total mass $3.4^{+0.3}_{-0.1}~M_\odot$. The mass ranges of components are from $1.12$ to $2.52 ~M_\odot$ with high-spin priors~\citep{abbott2020a}. Several months later, a new event of a compact binary merger with a $22.2-24.3 ~M_\odot$ black hole and a compact component with a mass of $2.50-2.67~M_\odot$ was reported by LVC as GW190814. The secondary object of GW190814 attracts a lot of attentions, since it may be either  the heaviest neutron star or the lightest black hole ever discovered~\citep{abbott2020b}.
		
		Since then, many interesting works were proposed to explain the secondary object of GW190814 in the past several months. Tan et al. considered that it may be a heavy neutron star including the deconfined QCD matter in the core~\citep{tan2020}. The possibility of a super-fast pulsar was assumed by Zhang et al.~\citep{zhangli2020} and Tsokaros et al.~\citep{tsokaros2020}. Bayesian modeling supported the neutron star with $2.50-2.60~M_\odot$ under the constraints on the properties of $1.4 ~M_\odot$ neutron star~\citep{lim2020}. On the other hand, it was also concluded that GW190814 may be a binary black hole merger by  Fattoyev et al.~\citep{fattoyev2020} and Tews et al.~\citep{tews2020}.
		
		The mass, radius and tidal deformabilities of neutron star are mainly determined by the equation of state (EOS) of neutron star matter, i. e., the relation between energy density and pressure. Many attempts have been made to obtain the EOS of supranuclear matter in neutron star from different models. It can be assumed as a simple polynomial in terms of pressure and energy density~\citep{annala2018}. It also can be generated by the nuclear density functional theories (DFT) \citep{vautherin1972, shen1998,  douchin2001,shen2002,long2006, long2007, sun2008,dutra2012, bao2014a,bao2014b}, where the nucleon-nucleon ($NN$) interaction was effectively determined by fitting the ground state properties of finite nuclei or the empirical saturation properties of infinite nuclear matter. Moreover, {\it ab initio} methods with realistic nuclear potentials extracted from $NN$ scattering are available to study the neutron star~\citep{akmal98,li06,carlson15,hu17,sammarruca18,logoteta19, wang2020}.
		
		In these available nuclear many-body models, the EOSs of nuclear matter around the saturation density ($\sim \rho_{B0}$) are well constrained, which corresponds to the central density of finite nuclei~\citep{dutra2012,dutra2014}. When these EOSs are  extrapolated to the supranuclear matter  ($\sim 5\rho_{B0}$), most of them can reasonably describe the properties of massive neutron star around $2.0~M_\odot$. There are only very few EOSs from the covariant density functional theory (CDFT) , which can generate the mass of neutron star above $2.5 ~M_\odot$ such as NL3 parameter set~\citep{lalazissis97}. However the radius of  $1.4 ~M_\odot$ from NL3 is too large to satisfy the recent constraints from the observations of GW170817 and NICER. Therefore, a new parameter set, BigApple, was proposed to generate a $2.6~M_\odot$ neutron star and reproduce the observables of finite nuclei and NICER~\citep{fattoyev2020}.
		
	    The CDFT has achieved great successes in the fields of nuclear physics and astrophysics. The first available version of CDFT was proposed by Walecka with the Hartree approximation, i. e., the $\sigma-\omega$ model~\citep{walecka1974}, which is also called as relativistic mean-field (RMF) model. Then, the $\rho$ meson, nonliear terms of $\sigma$ and $\omega$ mesons, and the coupling terms with $\rho$ meson to $\sigma$ or $\omega$ meson were introduced step by step in the RMF model~\citep{serot1979, boguta1977,sugahara1994, horowitz2001}. These nonlinear RMF models can describe the ground state properties of most nuclei in the nuclide chart precisely. Meanwhile, the contributions of the exchange terms in the mean-field approximation were considered at the end of 1970s in the framework of relativistic Hartree-Fock  (RHF) model, where the pion effect can be taken into account~\citep{brockmann1978,bouyssy1987,long2006,long2007}. The picture of meson exchange can be simplified as a point contact interaction when the meson masses are assumed to have an infinite value, which avoids solving the equation of motion for mesons~\citep{nikolauset1992}. This point coupling RMF model is also widely applied to study the nuclear mass table~\citep{zhao2010}. Furthermore, the nonlinear terms of various mesons could be replaced by the density-dependent meson-nucleon coupling constants in the density-dependent RMF (DDRMF) and DDRHF models, which consider the nuclear medium effect originated by the relativistic Brueckner-Hartree-Fock model~\citep{brockmann1992}. 
	
		Ten years ago, it was showed by Sun et al. ~\citep{sun2008} that some parameterizations of DDRMF and DDRHF models generated massive neutron stars around $2.33-2.48 ~M_\odot$ such as PKDD~\citep{long2004}, DD-ME1~\citep{niksic2002}, DD-ME2~\citep{lalazissis2005}, PKO1, PKO2, and PKO3~\citep{long2006} sets, whereas properties of neutron star at $1.4 ~M_\odot$ were not carefully discussed due to the  deficiencies of astronomical observables. In 2020, several DDRMF parameters, DD-MEX~\citep{taninah2020}, DD-LZ1~\citep{wei2020}, and DDV, DDVT, DDVTD~\citep{typel2020} were proposed by different groups by fitting ground state properties of spherical finite nuclei, which considered the parametric correlations, shell evaluations, and tensor couplings of the vector mesons to nucleons, respectively. Therefore, it is necessary to systematically calculate the properties of neutron star with these latest DDRMF parameterizations and discuss the possibility of the secondary object of GW190817 as a neutron star.
		
		This paper is arranged as follows: the theoretical descriptions of DDRMF model and neutron star matter are shown in Sec.~\ref{theorfram}; in Sec.~\ref{results}, properties of nuclear matter and neutron star will be presented and discussed with various DDRMF models.  The summaries and  discussion will be given in Sec.~\ref{summary}.
 
\section{The density-dependent relativistic mean-field model in neutron star}\label{theorfram}

In DDRMF model, the nucleons usually interact  with each other in nuclear system through exchanging the scalar-isoscalar ($\sigma$), vector-isoscalar ($\omega$), and  vector-isovector($\rho$) mesons. In some models, the scalar-isoscalar ($\delta$) meson is also taken into account to consider the isovector effect on the scalar potential of nucleon. The DDRMF Lagrangian density can be written as:
\begin{align}
\mathcal{L}_{DD}=&\sum_{i=p,~n}\overline{\psi}_i\left[\gamma^{\mu}\left(i\partial_{\mu}-\Gamma_{\omega}(\rho_B)\omega_{\mu}-\frac{\Gamma_{\rho}(\rho_B)}{2}\gamma^{\mu}\vec{\rho}_{\mu}\vec{\tau}\right)-\left(M-\Gamma_{\sigma}(\rho_B)\sigma-\Gamma_{\delta}(\rho_B)\vec{\delta}\vec{\tau}\right)\right]\psi_i\nonumber\\
&+\frac{1}{2}\left(\partial^{\mu}\sigma\partial_{\mu}\sigma-m_{\sigma}^2\sigma^2\right)+\frac{1}{2}\left(\partial^{\mu}\vec{\delta}\partial_{\mu}\vec{\delta}-m_{\delta}^2\vec{\delta}^2\right)\nonumber\\
&-\frac{1}{4}W^{\mu\nu}W_{\mu\nu}+\frac{1}{2}m_{\omega}^2\omega_{\mu}\omega^{\mu}-\frac{1}{4}\vec{R}^{\mu\nu}\vec{R}_{\mu\nu}+\frac{1}{2}m_{\rho}^2\vec{\rho}_{\mu}\vec{\rho}^{\mu},
\end{align}
where, $\psi_i$ represents the wave function of nucleon (proton or neutron). $\sigma,~\omega_{\mu},~\vec{\rho}_\mu$, and $\vec{\delta}$ denote the $\sigma,~\omega, ~\rho$, and $\delta$ mesons, respectively. $W_{\mu\nu}$ and $\vec{R}_{\mu\nu}$ are the anti-symmetry tensor fields of $\omega$ and $\rho$ mesons.
In nuclear matter, the tensor coupling between the vector meson and nucleon does not provide any contributions. Therefore, it is neglected in the present Lagrangian.  The coupling constants between mesons and nucleon are density-dependent in DDRMF model, which was firstly proposed by Brockmann and Toki~\citep{brockmann1992}. It takes into account that the $NN$ interaction in dense matter is affected by nuclear medium.  The density-dependent behaviors of the coupling constants have many styles. In CDFT, there are two types of density, i. e., the scalar density ($\rho_s$) and vector density ($\rho_B$). In principle, the coupling constants in DDRMF can be dependent on scalar density or vector density. In this work, we focus on the parameterizations of DDRMF depending on the vector density, which only influences the self energy instead of total energy.  Coupling constants of $\sigma$ and $\omega$ mesons are usually expressed as a fraction function of the vector density.  In DD2~\citep{niksic2002}, DD-ME1, DD-ME2, DDME-X, DDV, DDVT, and DDVTD parameterizations, they are given as:
\begin{equation}\label{cdd}
\Gamma_i(\rho_B)=\Gamma_i(\rho_{B0})f_i(x),~~\text{with}~~f_i(x)=a_i\frac{1+b_i(x+d_i)^2}{1+c_i(x+d_i)^2},~x=\rho_B/\rho_{B0},
\end{equation}
for $i=\sigma,~\omega$. $\rho_{B0}$ is the saturation density of symmetric nuclear matter. Five constraints on the coupling constants $f_i(1)=1,~f_i^{''}(0)=0,~f_{\sigma}^{''}(1)=f_{\omega}^{''}(1)$ can reduce the numbers of independent parameters to three in Eq.~\eqref{cdd}. The first two constraints lead to
\begin{gather}
a_i=\frac{1+c_i(1+d_i)^2}{1+b_i(1+d_i)^2},~~~~3c_id_i^2=1.
\end{gather}
For the isovector mesons $\rho$ and $\delta$, their coupling constants are,
\begin{gather}
\Gamma_i(\rho_B)=\Gamma_i(\rho_{B0}){\rm exp}[-a_i(x-1)].
\end{gather}
While in DD-LZ1 parametrization, the coefficient in front of fraction function, $\Gamma_i$ is fixed at $\rho_B=0$ for $i=\sigma,~\omega$:
\begin{gather}\label{1.5}
\Gamma_i(\rho_B)=\Gamma_i(0)f_i(x),
\end{gather}
There are only four constraint conditions as $f_i(0)=1$ and $f''_i(0)=0$ for $\sigma$ and $\omega$ coupling constants in DD-LZ1. The constraint $f''_{\sigma}(1)=f''_{\omega}(1)$ is removed in DD-LZ1 parametrization, which can give more precise shell evaluations of finite nuclei around $Z=58$ and $92$~\citep{wei2020}. For $\rho$ meson, its coupling constant is also changed accordingly as 
\begin{gather}\label{1.6}
\Gamma_{\rho}(\rho_B)=\Gamma_{\rho}(0){\rm exp}(-a_{\rho}x).
\end{gather}

To solve the nuclear many-body system in the DDRMF model, the mean-field approximation must be adopted following the nonlinear RMF models, in which various mesons are treated as classical fields as
\begin{gather}
\sigma~\rightarrow\left\langle\sigma\right\rangle\equiv\sigma,~\omega_{\mu}\rightarrow\left\langle\omega_{\mu}\right\rangle\equiv\omega,
~\vec{\rho}_{\mu}\rightarrow\left\langle\vec{\rho}_{\mu}\right\rangle\equiv\rho,~\vec{\delta}\rightarrow\left\langle\vec{\delta}\right\rangle\equiv\delta,~\langle\psi\rangle\rightarrow\psi.
\end{gather}
The space components of vector meson are removed in the  parity conservation system. In addition, the spatial derivatives about nucleon and mesons are neglected in the infinite nuclear matter due to its transformation invariance. Finally, using the Euler-Lagrange equation, the equations of motion of nucleon and mesons are obtained:
\begin{align}
\label{1.8}
 &\sum_{i=p,n}\left[i\gamma^{\mu}\partial_{\mu}-\gamma^0\left(\Gamma_{\omega}(\rho_B)\omega+\frac{\Gamma_{\rho}(\rho_B)}{2}\rho\tau_3+\Sigma_R(\rho_B)\right)-M_i^{*}\right]\psi_i=0.\nonumber\\
&m_{\sigma}^2\sigma=\Gamma_{\sigma}(\rho_B)\rho_s,\nonumber\\
&m_{\omega}^2\omega=\Gamma_{\omega}(\rho_B)\rho_B,\nonumber\\
&m_{\rho}^2\rho=\frac{\Gamma_{\rho}(\rho_B)}{2}\rho_3,\nonumber\\
&m_{\delta}^2\delta=\Gamma_{\delta}(\rho_B)\rho_{s3}.
\end{align}
The isospin third components of nucleon are defined as $\tau_3=1$ and $\tau_3=-1$ for protons and neutrons, respectively. A rearrangement term $\Sigma_R$ will be introduced into Eq.~\eqref{1.8} due to the density dependence of coupling constants,
\begin{gather}
\Sigma_R(\rho_B)=-\frac{\partial\Gamma_{\sigma}(\rho_B)}{\partial\rho_B}\sigma\rho_s-\frac{\partial\Gamma_{\delta}(\rho_B)}{\partial\rho_B}\delta\rho_{s3}+\frac{\partial\Gamma_{\omega}(\rho_B)}{\partial\rho_B}\omega\rho_B+\frac{1}{2}\frac{\partial\Gamma_{\rho}(\rho_B)}{\partial\rho_B}\rho\rho_3,
\end{gather}
where the scalar, vector densities, and their isospin components are generated by the expectation value of nucleon fields,
\begin{align}
\rho_s=\left\langle\overline{\psi}\psi\right\rangle=\rho_{sp}+\rho_{sn},~~~~
&\rho_{s3}=\left\langle\overline{\psi}\tau_3\psi\right\rangle=\rho_{sp}-\rho_{sn},\nonumber\\
\rho_{B}=\left\langle\psi^{\dag}\psi\right\rangle=\rho_{Bp}+\rho_{Bn},~~~~
&\rho_3=\left\langle\psi^{\dag}\tau_3\psi\right\rangle=\rho_{Bp}-\rho_{Bn}.
\end{align}

The effective masses of nucleons in Eq.~\eqref{1.8} are dependent on the scalar mesons $\sigma$ and $\delta$,
\begin{align}
M_p^{*}&=M-\Gamma_{\sigma}(\rho_B)\sigma-\Gamma_{\delta}(\rho_B)\delta,\nonumber\\
M_n^{*}&=M-\Gamma_{\sigma}(\rho_B)\sigma+\Gamma_{\delta}(\rho_B)\delta
\end{align}
and the corresponding effective energies of nucleons have the following form because of the mass-energy relation,
\begin{gather}
E_{Fi}^{*}=\sqrt{k_{Fi}^2+(M_{i}^{*})^2},
\end{gather}
where $k_{Fi}$ is the Fermi momentum of nucleon.

With the energy-momentum tensor in a uniform system, the energy density, $\mathcal{E}$ and pressure, $P$ of infinite nuclear matter can be obtained respectively as
\begin{align}
\mathcal{E}_{\rm DD}=&\frac{1}{2}m_{\sigma}^2\sigma^2-\frac{1}{2}m_{\omega}^2\omega^2-\frac{1}{2}m_{\rho}^2\rho^2+\frac{1}{2}m_{\delta}^2\delta^2
+\Gamma_{\omega}(\rho_B)\omega\rho_B+\frac{\Gamma_{\rho}(\rho_B)}{2}\rho\rho_3+\mathcal{E}_{\rm kin}^p+\mathcal{E}_{\rm kin}^n,\\\nonumber
P_{\rm DD}=&\rho_B\Sigma_{R}(\rho_B)-\frac{1}{2}m_{\sigma}^2\sigma^2+\frac{1}{2}m_{\omega}^2\omega^2+\frac{1}{2}m_{\rho}^2\rho^2-\frac{1}{2}m_{\delta}^2\delta^2
+P_{\rm kin}^p+P_{\rm kin}^n,
\end{align}
where, the contributions from kinetic energy are 
\begin{align}
\mathcal{E}_{\rm kin}^i&=\frac{\gamma}{2\pi^2}\int_{0}^{k_{Fi}}k^2\sqrt{k^2+{M_i^{*}}^{2}}dk=\frac{\gamma}{16\pi^2}\left[k_{Fi}E_{Fi}^{*}\left(2k_{Fi}^2+{M_i^{*}}^2\right)+{M_i^{*}}^4{\rm ln}\frac{M_i^{*}}{k_{Fi}+E_{Fi}^{*}}\right], \\\nonumber
P_{\rm kin}^i&=\frac{\gamma}{6\pi^2}\int_{0}^{k_{Fi}}\frac{k^4 dk}{\sqrt{k^2+{M_i^{*}}^{2}}}
=\frac{\gamma}{48\pi^2}\left[k_{Fi}\left(2k_{Fi}^2-3{M_i^{*}}^2\right)E_{Fi}^{*}+3{M_i^{*}}^{4}{\rm ln}\frac{k_{Fi}+E_{Fi}^{*}}{M_i^{*}}\right].
\end{align}
$\gamma=2$ is the spin degeneracy factor. The binding energy per nucleon can be defined by
\begin{align}
\frac{E}{A}&=\frac{\mathcal{E}}{\rho_{B}}-M.
\end{align}
The symmetry energy is calculated by
\begin{align}
E_{\rm symDD}=\frac{1}{2}\frac{\partial^2E/A}{\partial\beta^2},
\end{align}
where $\beta$ is the asymmetry factor, defined as $\beta=(\rho_{Bn}-\rho_{Bp})/(\rho_{Bn}+\rho_{Bp})$ and its slope, $L_{DD}$ is given by
\begin{gather}
L_{DD}=3\rho_{B}\frac{\partial E_{\rm symDD}}{\partial\rho_B}\bigg|_{\rho_B=\rho_{B0}}.
\end{gather}
Actually, both of them can be derived analytically in RMF model~\citep{dutra2014}.

The scalar potential and vector potential of nucleon are expressed as,
\begin{gather}
U_S=\Gamma_{\sigma}(\rho_B)\sigma+\Gamma_{\delta}(\rho_B)\delta\tau_3,\\
U_V=\Gamma_{\omega}(\rho_B)\omega+\frac{1}{2}\Gamma_{\rho}(\rho_B)\rho\tau_3
+\left[-\frac{\partial\Gamma_{\sigma}(\rho_B)}{\partial\rho_B}\sigma\rho_s-\frac{\partial\Gamma_{\delta}(\rho_B)}{\partial\rho_B}\delta\rho_{s3}+\frac{\partial\Gamma_{\omega}(\rho_B)}{\partial\rho_B}\omega\rho_B+\frac{1}{2}\frac{\partial\Gamma_{\rho}(\rho_B)}{\partial\rho_B}\rho\rho_3\right],
\end{gather}
where the derivative terms in the vector potential originate from the density dependence of coupling constants.

The outer core part of a neutron star, which almost dominates its mass and radius, is usually treated as the uniform matter composed of neutron, proton, and leptons. They are stably existing with the conditions of beta equilibrium and charge neutrality.  Therefore the chemical potentials of nucleons and leptons are very important, that can be derived from the thermodynamics equations at zero temperature, 
\begin{gather}\label{2.1}
\mu_{Bi}=\sqrt{k_{Fi}^2+M_i^{*2}}+\left[\Gamma_{\omega}(\rho_B)\omega+\frac{\Gamma_{\rho}(\rho_B)}{2}\rho\tau_3+\Sigma_R(\rho_B)\right],\\\nonumber
\label{2.2}
\mu_l=\sqrt{k_{Fl}^2+m_l^{2}}.
\end{gather}
In neutron star matter, with the density increasing, the muon will be onset when the electron chemical potential $\mu_e$ is larger than the muon rest mass, i. e., $\mu_e>m_{\mu}=106.55$ MeV. Hence, the beta equilibrium condition now can be expressed by
\begin{gather}\label{2.3}
\mu_{\mu}=\mu_e=\mu_n-\mu_p.
\end{gather}
The charge neutrality condition has the following form:
\begin{gather}\label{2.4}
\rho_{Bp}=\rho_e+\rho_{\mu}.
\end{gather}

The pressure and energy density will be obtained as a function of nucleon density within the constraints of Eqs.~\eqref{2.3} and \eqref{2.4}. The Tolman-Oppenheimer-Volkoff (TOV) equation\citep{tolman1939,oppenheimer1939}  describes a spherically symmetric star in the gravitational equilibrium from general relativity,
\begin{gather}\label{tov}
\frac{dP}{dr}=-\frac{GM(r)\mathcal{E}(r)}{r^2}\frac{\left[1+\frac{P(r)}{\mathcal{E}(r)}\right]\left[1+\frac{4\pi r^3P(r)}{M(r)}\right]}{1-\frac{2GM(r)}{r}},\\\nonumber
\frac{dM(r)}{dr}=4\pi r^2\mathcal{E}(r),
\end{gather}
where $P$ and $M$ are the pressure and mass of neutron star at $r$. Furthermore, the tidal deformability becomes a typical property of neutron star after the observation of the gravitational wave from BNS merger, which characterizes the deformation of a compact object in an external field generated by another star.  The tidal deformability of a neutron star can be reduced as dimensionless form,
\begin{gather}\label{dt}
\Lambda=\frac{2}{3}k_2C^{-5}.
\end{gather}
where $C=GM/R$ is the compactness parameter. The second order Love number $k_2$~\citep{hinderer2008,hinderer2010} is given by
\begin{align}\label{lneq}
k_2=&\frac{8C^5}{5}(1-2C)^2\left[2+2C(\mathcal{Y}_R-1)-\mathcal{Y}_R\right]\Big\{2C\left[6-3\mathcal{Y}_R+3C(5\mathcal{Y}_R-8)\right]\nonumber\\
&+4C^3\left[13-11\mathcal{Y}_R+C(3\mathcal{Y}_R-2)+2C^2(1+\mathcal{Y}_R)\right]\nonumber\\
&+3(1-2C)^2\left[2-\mathcal{Y}_R+2C(\mathcal{Y}_R-1){\rm ln}(1-2C)\right]\Big\}^{-1}.
\end{align}
Here, $\mathcal{Y}_R=y(R)$. $y(r)$ satisfies the following first-order differential equation,
\begin{equation}
r\frac{d y(r)}{dr} + y^2(r)+y(r)F(r) + r^2Q(r)=0,
\end{equation}
where $F(r)$ and $Q(r)$ are functions related to the pressure and energy density
\begin{align}
F(r) & = \left[1-\frac{2M(r)}{r}\right]^{-1} 
\left\{1-4\pi r^2[\mathcal{E}(r)-P(r)]\right\} ,\\ 
\nonumber 
r^2Q(r) & =  \left\{4\pi r^2 \left[5\mathcal{E}(r)+9P(r)+\frac{\mathcal{E}(r)
	+P(r)}{\frac{\partial P}{\partial \mathcal{E}}(r)}\right]-6\right\}\times 
\left[1-\frac{2M(r)}{r}\right]^{-1}  \\\nonumber
&~~-\left[\frac{2M(r)}{r} +2\times4\pi r^2 P(r) \right]^2 \times 
\left[1-\frac{2M(r)}{r}\right]^{-2} .
\end{align}
The second Love number corresponds to the initial condition $y(0)=2$. It is also related to the speed of sound in compact matter, $c_s$
\begin{gather}
c_s^2=\frac{\partial P(\varepsilon)}{\partial{\mathcal{E}}}.
\end{gather}

\section{The results and discussions}\label{results}
Firstly,  masses of nucleons and mesons, coupling constants between nucleon and mesons, and saturation densities of symmetric nuclear matter,  $\rho_{B0}$ in DD2~\citep{typel2010}, DD-ME1~\citep{niksic2002}, DD-ME2~\citep{lalazissis2005}, DDME-X~\citep{taninah2020}, DDV, DDVT, DDVTD~\citep{typel2020}, and DD-LZ1~\citep{wei2020} sets are all listed in Table~\ref{table.1},
 \begin{table}[htb]
 	\centering
 	\caption{ Masses of nucleons and mesons, meson coupling constants, and the nuclear saturation densities in various DDRMF models.}\label{table.1}
 	\scalebox{0.8}{
 	\begin{tabular}{lrrrrrrrrrrr}
 		\hline\hline
 		&DD-LZ1    ~~& ~~&DD2 ~~&DD-ME1~~&DD-ME2~~&DD-MEX~~&DDV\      ~~&DDVT  ~~&DDVTD\\
 		\hline
 		$m_n[\rm MeV]$    &938.900000  ~~&$m_n$  ~~&939.56536 ~~&939.0000   ~~&939.0000   ~~&939.0000 ~~&939.565413~~&939.565413 ~~&939.565413\\
 		$m_p[\rm MeV]$    &938.900000  ~~&$m_p$  ~~&938.27203 ~~&939.0000   ~~&939.0000   ~~&939.0000 ~~&938.272081~~&938.272081~~&938.272081 \\
 		$m_{\sigma}[\rm MeV]$ &538.619216  ~~&$m_{\sigma}$  ~~&546.212459 ~~&549.5255 ~~&550.1238 ~~&547.3327  ~~&537.600098~~&502.598602~~&502.619843\\
 		$m_{\omega}[\rm MeV]$ &783.0000    ~~&$m_{\omega}$  ~~&783.0000   ~~&783.0000   ~~&783.0000   ~~&783.0000  ~~&783.0000  ~~&783.0000  ~~&783.0000   \\
 		$m_{\rho}[\rm MeV]$   &769.0000    ~~&$m_{\rho}$     ~~&763.0000   ~~&763.0000   ~~&763.0000   ~~&763.0000  ~~&763.0000  ~~&763.0000  ~~&763.0000   \\
 		$m_{\delta}[\rm MeV]$ &---         ~~&$m_{\delta}$    ~~&---      ~~&---      ~~&---        ~~&---       ~~&---       ~~&---       ~~&980.0000\\
 		$\Gamma_{\sigma}(0)$ &12.001429~~&$\Gamma_{\sigma}(\rho_{B0})$  ~~&10.686681  ~~&10.4434    ~~&10.5396    ~~&10.7067   ~~&10.136960 ~~&8.382863  ~~&8.379269   \\
 		$\Gamma_{\omega}(0)$ &14.292525~~&$\Gamma_{\omega}(\rho_{B0})$  ~~&13.342362  ~~&12.8939    ~~&13.0189    ~~&13.3388   ~~&12.770450 ~~&10.987106 ~~&10.980433  \\
 		$\Gamma_{\rho}(0)$  &15.150934 ~~&$\Gamma_{\rho}(\rho_{B0})$  ~~&7.25388    ~~&7.6106     ~~&7.3672     ~~&7.2380    ~~&7.84833   ~~&7.697112  ~~&8.06038    \\
 		$\Gamma_{\delta}(0)$ &---      ~~&$\Gamma_{\delta}(\rho_{B0})$ ~~&---        ~~&---        ~~&---        ~~&---       ~~&---   ~~&---       ~~&0.8487420  \\
 		\hline
 		$\rho_{B0}[\rm fm^{-3}]$ &0.158100 ~~&$\rho_{B0}$    ~~&0.149    ~~&0.152      ~~&0.152      ~~&0.153     ~~&0.1511    ~~&0.1536    ~~&0.1536\\
 		\hline
 		$a_{\sigma}$  &1.062748 ~~&$a_{\sigma}$  ~~&1.357630  ~~&1.3854  ~~&1.3881  ~~&1.3970  ~~&1.20993    ~~&1.20397    ~~&1.19643    \\
 		$b_{\sigma}$  &1.763627 ~~&$b_{\sigma}$  ~~&0.634442  ~~&0.9781  ~~&1.0943  ~~&1.3350  ~~&0.21286844 ~~&0.19210314 ~~&0.19171263 \\
 		$c_{\sigma}$  &2.308928 ~~&$c_{\sigma}$  ~~&1.005358  ~~&1.5342  ~~&1.7057   ~~&2.0671    ~~&0.30798197~~&0.27773566~~&0.27376859 \\
 		$d_{\sigma}$  &0.379957 ~~&$d_{\sigma}$  ~~&0.575810  ~~&0.4661  ~~&0.4421     ~~&0.4016    ~~&1.04034342~~&1.09552817~~&1.10343705 \\
 		$a_{\omega}$  &1.059181 ~~&$a_{\omega}$  ~~&1.369718  ~~&1.3879  ~~&1.3892     ~~&1.3936    ~~&1.23746   ~~&1.16084   ~~&1.16693    \\
 		$b_{\omega}$  &0.418273 ~~&$b_{\omega}$  ~~&0.496475   ~~&0.8525     ~~&0.9240  ~~&1.0191    ~~&0.03911422~~&0.04459850~~&0.02640016 \\
 		$c_{\omega}$  &0.538663 ~~&$c_{\omega}$  ~~&0.817753   ~~&1.3566     ~~&1.4620     ~~&1.6060    ~~&0.07239939~~&0.06721759~~&0.04233010 \\
 		$d_{\omega}$  &0.786649 ~~&$d_{\omega}$  ~~&0.638452   ~~&0.4957     ~~&0.4775     ~~&0.4556    ~~&2.14571442~~&2.22688558~~&2.80617483 \\
 		$a_{\rho}$    &0.776095 ~~&$a_{\rho}$    ~~&0.518903   ~~&0.5008     ~~&0.5647     ~~&0.6202    ~~&0.35265899~~&0.54870200~~&0.55795902 \\
 		$a_{\delta}$  &---   ~~&$a_{\delta}$   ~~&---        ~~&---        ~~&---        ~~&---       ~~&---       ~~&---       ~~&0.55795902 \\
 		\hline\hline
 	\end{tabular}}
 \end{table}

The mass of $\sigma$ meson is fitted as a free parameter in DDRMF model. The coefficients of meson coupling constants, $\Gamma_i$  in DD-LZ1 are the values at zero density, while other parameter sets adopted the values at nuclear saturation densities. The magnitudes of $\Gamma_\sigma(\rho_{B0}), ~\Gamma_\omega(\rho_{B0})$ and $\Gamma_\rho(\rho_{B0})$ in DD2, DDME-1, DD-ME2, DD-MEX, DDV are consistent with each other. The tensor couplings between  vector mesons and nucleon were considered in DDVT and DDVTD, where $\Gamma_\sigma(\rho_{B0})$ and $\Gamma_\omega(\rho_{B0})$ have significant differences comparing to other parameter sets. In addition, the $\delta$ meson is included in DDVTD set. 
 \begin{figure}[htb]
 	\centering
 	\includegraphics[scale=0.4]{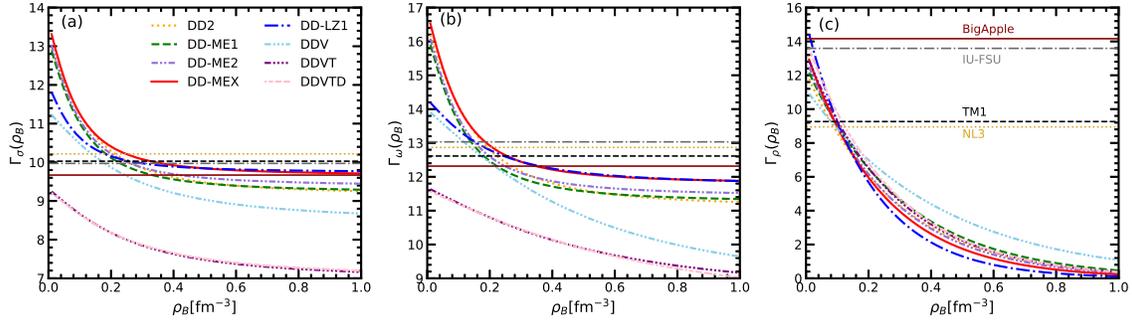}
 	\caption{The coupling constants of $\omega,~\sigma$, and $\rho$ mesons as functions of vector density  in various DDRMF models and several nonlinear RMF models.}\label{fig.1}
 	\end{figure}
 
 To show the density-dependent behaviors of these coupling constants more clearly, they are plotted as functions of the vector density in Fig.~\ref{fig.1}. It can be found that all of these coupling constants decrease when the nuclear density becomes larger due to the nuclear medium effect. For the $\rho$ meson coupling constants in panel (c), all parameter sets have very similar density-dependent behaviors in the whole density region. In DDVT and DDVTD, the tensor coupling constants play obvious roles in finite nuclei due to their derivative forms, however, they do not provide any contribution in nuclear matter. Their coupling constants of $\sigma$ and $\omega$ mesons in panel (a) and panel (b) are dramatically smaller than other sets. Furthermore, the coupling constants from several typical nonlinear RMF models, NL3~\citep{lalazissis97}, TM1~\citep{sugahara1994}, IUFSU~\citep{horowitz2001}, and BigApple~\citep{fattoyev2020} are also shown to compare their differences with those in DDRMF model. At low density region, the coupling constants in DDRMF models are usually stronger than those in nonlinear RMF modes, while weaker at higher density.
 
 With these DDRMF parameter sets, the saturation properties of nuclear matter can be calculated, such as the saturation density, binding energy, incompressibility, symmetry energy, the slope of symmetry energy, and the effective nucleon mass.  In Table~\ref{table.2}, these properties calculated by various DDRMF models are listed, whose uncertainties of different parameter sets are very small in saturation density, binding energy, incompressibility, and symmetry energy. The slopes of symmetry energy from different models, $L$ are around $40-70$ MeV, which also satisfy the recent constraints, $L=59.57\pm10.06$ MeV \citep{zhang2020}. On the other hand, the effective nucleon masses in DDVT and DDVTD are relatively larger, since their scalar coupling strengths are much smaller comparing to other sets.
 \begin{table}[htb]
 	\footnotesize
 	\centering
 	\caption{Nuclear matter properties at saturation density generated by present DDRMF parameterizations.}\label{table.2}
 	\begin{tabular}{lrrrrrrrrrrr}
 		\hline\hline
 		&DD-LZ1   ~~&DD2       ~~&DD-ME1     ~~&DD-ME2     ~~&DD-MEX    ~~&DDV       ~~&DDVT        ~~&DDVTD  \\
 		\hline
 		$\rho_{B0}[\rm fm^{-3}]$   &0.1585   ~~&0.149     ~~&0.152      ~~&0.152      ~~&0.1518    ~~&0.1511    ~~&0.1536     ~~&0.1536   \\
 		$E/A[\rm MeV]$                  &-16.126  ~~&-16.916   ~~&-16.668    ~~&-16.233    ~~&-16.14    ~~&-16.097  ~~&-16.924    ~~&-16.915    \\
 		$K_0[\rm MeV]$                  &231.237  ~~&241.990   ~~&243.881    ~~&251.306    ~~&267.059   ~~&239.499   ~~&239.999    ~~&239.914\\
 		$E_{\rm sym}[\rm MeV]$          &32.016   ~~&31.635    ~~&33.060     ~~&32.31      ~~&32.269    ~~&33.589    ~~&31.558     ~~&31.817  \\
 		$L[\rm MeV]$                  &42.467   ~~&54.933    ~~&55.428     ~~&51.265     ~~&49.692    ~~&69.646    ~~&42.348     ~~&42.583 \\
 		$M_n^{*}/M$                     &0.558   ~~&0.563    ~~&0.578     ~~&0.572      ~~&0.556    ~~&0.586    ~~&0.667     ~~&0.667\\
 		$M_p^{*}/M$                     &0.558   ~~&0.562    ~~&0.578     ~~&0.572      ~~&0.556    ~~&0.585    ~~&0.666     ~~&0.666 \\
 		\hline\hline
 	\end{tabular}
 \end{table}
 
\begin{figure}[htb]
 	\centering
 	%\hspace*{-15pt}
 	\includegraphics[scale=0.6]{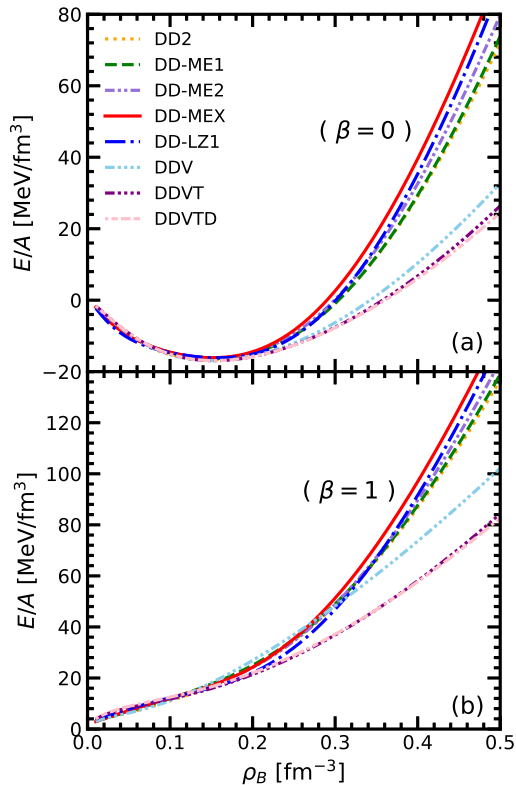}
 	\caption{Equations of state of symmetric nuclear matter($\beta=0$) in panel (a) and pure neutron matter($\beta=1$) in panel(b) from various DDRMF models.}\label{fig.2}
 \end{figure}

The binding energies per nucleon for symmetric nuclear matter in panel (a) of Fig~.\ref{fig.2} and pure neutron matter in panel (b) of Fig~.\ref{fig.2} as functions of vector density are plotted with the present DDRMF parameterizations. These equations of state (EOSs) of nuclear matter below $0.2$ fm$^{-3}$ are almost identical since all the parameters were determined by properties of finite nuclei, whose central density is around nuclear saturation density $\rho_{B0}\sim0.15$ fm$^{-3}$. Their differences increase from $0.30$ fm$^{-3}$. In symmetric nuclear matter, they are separated into the softer group with DDV, DDVT, and DDVTD, and the stiffer group with DD2, DD-ME1, DD-ME2, DD-MEX, and DD-LZ1. The scalar and vector coupling strengths in softer group sets are obviously weaker than those in stiffer group sets. The binding energy of pure neutron matter from DDV is larger than the ones from DDVT and DDVTD. The DDV set has the largest slope of symmetry energy in the present DDRMF parameterizations. This slope will determine the density dependent behaviors of symmetry energy and the binding energy of pure neutron matter, due to $E/A(\beta=1)\approx E/A(\beta=0)+E_\text{sym}$ at a fixed density.

\begin{figure}[htb]
 	\centering
 	%\hspace*{-15pt}
 	\includegraphics[scale=0.6]{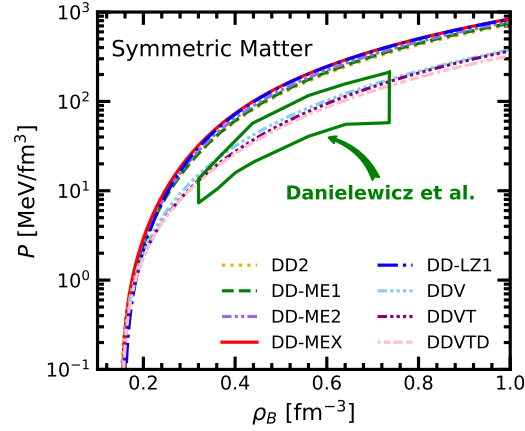}
 	\caption{Pressures as a function of vector density of symmetric nuclear matter with various DDRMF parameter sets and the constraints from the heavy ion collisions.}\label{fig.3}
 \end{figure}

In general, it is very difficult to measure properties of nuclear matter above twice nuclear saturation density from finite nuclei. Recently, the experiments about heavy-ion collisions provide us some useful information to constrain the EOS of nuclear matter at high density. In Fig.~\ref{fig.3}, the pressures in symmetric nuclear matter as functions of density from various DDRMF models are shown and compared to the constraints from heavy-ion collisions at $2-4\rho_{B0}$ by Danielewicz et al.~\citep{danielewicz2002}. We can find that the EOSs from the softer group sets are completely consistent with the experiment data, while the other group is indeed stiffer than the heavy-ion collisions constraints. We also notice that the BigApple and NL3 sets also have the similar situations in the work by Fattoyev et al.~\citep{fattoyev2020}. However, we want to emphasize here that the constraints from the heavy-ion collisions are strongly model-dependent, which is determined by many inputs, such as the $NN$ interaction. To our knowledge, there were few investigations about heavy-ion collisions, which adopted the RMF model as $NN$ interaction. Therefore, it cannot certainly claim that the EOSs generated by DD2, DD-ME1, DD-ME2, DD-MEX, and DD-LZ1 parameterizations are clearly excluded by the constraints of heavy-ion collisions.

\begin{figure}[htb]
 	\centering
 	\hspace*{-15pt}
 	\includegraphics[scale=0.6]{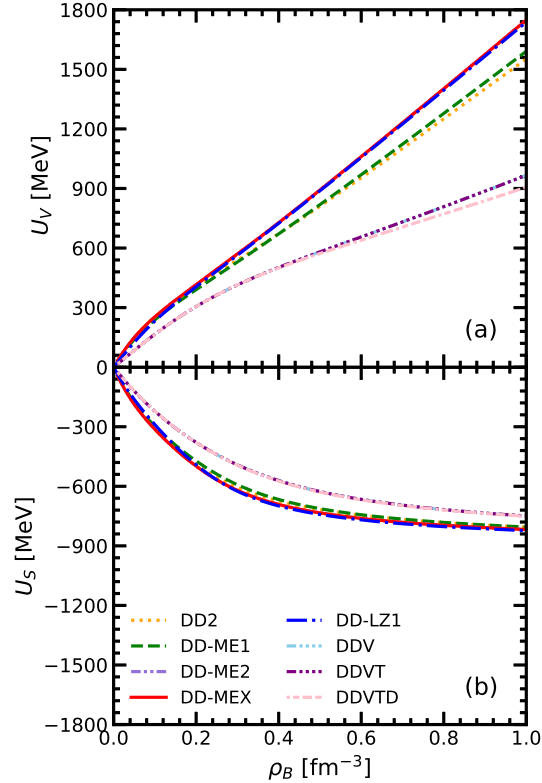}
 	\caption{Scalar and vector potentials as a function of the vector density from various DDRMF models.}\label{fig.4}
 \end{figure}

To explain the stiff EOSs at high density of  DD2, DD-ME1, DD-ME2, DD-MEX, and DD-LZ1 sets, the vector potentials in panel (a) of Fig.~\ref{fig.4}  and scalar potentials in panel (b) of Fig.~\ref{fig.4}  for symmetric nuclear matter from present DDRMF parameterizations are shown. The scalar potentials in these sets are very similar, while the vector potentials from different parameter sets have significant differences. The softer group sets provide the weakest vector potentials, which have the analogous magnitudes to the scalar potentials of nucleon. On the other hand, the DD-ME2, DD-MEX, and DD-LZ1 generate the strongest vector components, which are almost twice of those from DDV, DDVT, and DDVTD, because their $\omega$ coupling constants are largest at high density regions. Therefore, they provide very stiff EOSs.
 
\begin{figure}[htb]
	\centering
	\hspace*{-15pt}
	\includegraphics[scale=0.6]{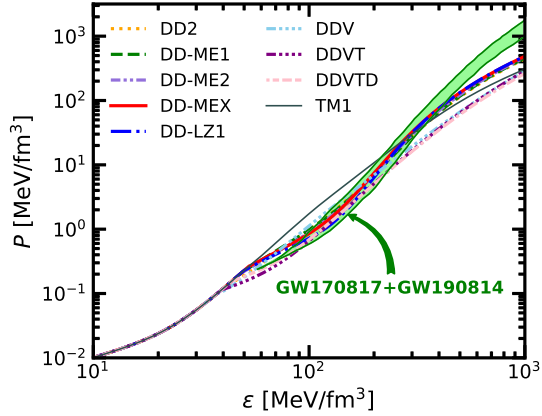}
	\caption{The pressure $P$ versus energy density $\varepsilon$ of neutron star matter from DDRMF models and joint constraints from GW170817 and GW190814. }\label{fig.5}
\end{figure}

Together with the conditions of beta equilibrium and charge neutrality, the EOSs of neutron star matter with DDRMF model can be obtained in Fig.~\ref{fig.5}, which shows the pressures of neutron star matter as a function of energy density. At crust part of neutron star, the EOS in the non-uniform matter generated by TM1 parametrization with Thomas-Fermi approximation is adopted. In the core of neutron star, EOSs of the uniform matter are calculated with various DDRMF sets. Their density-dependent behaviors are very similar with those in symmetric nuclear matter. At high density region, the stiffer group sets provide higher pressures due to the stronger vector potentials. The joint constraints on EOS extracted from the GW170817 and GW190814 are shown as a shaded band here. When the energy density is smaller than $600$ MeV/fm$^{3}$, the EOSs from stiffer group sets satisfy the constraints from the gravitational wave detection.  While, the pressures obtained from softer group sets start to be lower than the constraint band from $\varepsilon =300$ MeV/fm$^{3}$. Furthermore, the EOS from TM1 set is also given, which is stiffer than those from DDRMF at intermediate density region and becomes softer when density is creasing.  Since, the slope of symmetry energy in TM1 is around $110$ MeV. It is much larger than those derived from present DDRMF models, whose $L$ are around $40-70$ MeV. The slope of symmetry energy mainly influences the pressures in intermediate density. At higher density, the vector potentials in DDRMF from the stiffer group sets are stronger than the one in TM1.

\begin{figure}[htb]
	\centering
	\hspace*{-15pt}
	\includegraphics[scale=0.6]{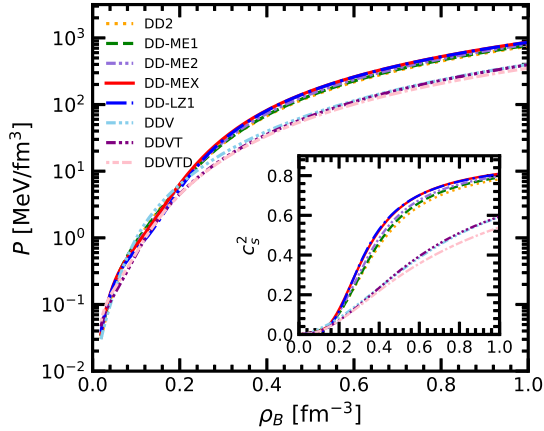}
	\caption{EOSs of neutron star matter with different DDRMF models.The corresponding speeds of sound in units of the speed of light shown in insert.}\label{fig.6}
\end{figure}

In Fig.~\ref{fig.6}, the pressures as functions of density in neutron star matter from DDRMF models are given. The pressures from the stiffer group sets are obviously larger than those generated by the softer group sets. The speeds of sound in neutron star matter, $c_s$ with the unit of light speed are plotted in the insert. The $c^2_s$ from softer group sets are much lower than those from other parameterizations, which are around $0.6$ at $\rho_B=1.0$ fm$^{-3}$. They are consistent with the results from nonlinear RMF models~\citep{hu2020}. The speed of  sound from stiffer group EOSs rapidly increase from $\rho_B=0.2$ fm$^{-3}$ and $c^2_s$ reach around $0.8$ at high density. They will be constants less than one as the density continues growing. Actually, the EOS and speed of light of BigApple set in nonlinear RMF model are very similar with the present work, where a $\omega-\rho$ coupling term was included to reduce the slope of symmetry energy and its vector-isovector coupling constant is very strong as we shown in Fig.~\ref{fig.1}~\citep{fattoyev2020}. 

\begin{figure}[htb]
	\centering
	\hspace*{-15pt}
	\includegraphics[scale=0.5]{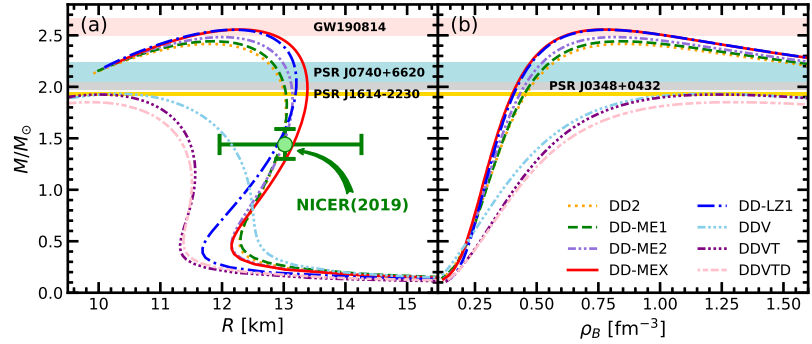}
	\caption{The neutron star masses as functions of  radius and the central baryon density. Constraints from astronomical observables for massive neutron star and NICER are also shown. }\label{fig.7}
\end{figure}

The mass-radius relation of a static neutron star can be solved by TOV equation Eq.~\eqref{tov}, where the EOS of neutron star matter is used. In Fig~\ref{fig.7}, the mass-radius ($M-R$) relation in panel (a) and mass-central density ($M-\rho_B$)  relation in panel (b) from various DDRMF models are shown.  From the panel (a), it can be found that the maximum masses of neutron star in softer group sets are around $1.85-1.93~M_\odot$ and the corresponding radii are $9.85-10.34$ km. These results only can explain the existence of  PSR J1614-2230 ($1.928\pm0.017~M_\odot$). As we discussed before, the EOSs from these three parameter sets are relatively soft due to their small vector potentials. The maximum masses calculated by DD2, DD-ME1, and DD-ME1 sets are about  $2.42-2.48~M_\odot$, which are consistent with the available investigations~\citep{sun2008}. DD-MEX and DD-LZ1 can support the neutron star above $2.5~M_\odot$ because of their strongest repulsive contributions from $\omega$ meson. Their maximum masses can approach $2.56~M_\odot$, which are in accord with the observed mass of the secondary compact object in GW190814, $2.50-2.67~M_\odot$. In addition to the constraints from the observables of  massive neutron stars, PSR J1614-2230, PSR J034+0432, and PSR J0740+6620, recently the mass and radius of neutron star  at intermediate mass region were measured simultaneously for PSR J0030+0451 by NICER.  Its mass and radius were reported around $1.4~M_\odot$ and $13$ km~\citep{miller2019,riley2019}. These constraints from NICER are plotted in panel (a). It can be found that the $M-R$ relations from stiffer group parameterizations around $1.4 ~M_\odot$ completely satisfy the observables from NICER, while the radii of neutron star at $1.4~M_\odot$, $R_{1.4}$ from DDVT, and DDVTD are around $11.4$ km, which are  smaller than the possible radii of J0030+0451.  The $R_{1.4}$ of DDV is $12.2$ km since its slope of symmetry energy is obviously larger than those of DDVT and DDVTD. When the isoscalar properties of RMF models are the same, the slope of symmetry energy can influence the radii of neutron star at $1.4 ~M_\odot$ in our recent investigations~\citep{ji2019,hu2020}.

The $M-\rho_B$ relations in panel (b) from present DDRMF models can be separated by two groups. The first group only can generate the neutron star around $1.9 ~M_\odot$ at the central densities $\sim8\rho_{B0}$ from softer group EOSs. The second group can support neutron stars around $2.5 ~M_\odot$, where the central densities locate at $5\rho_{B0}$. The corresponding speeds of sound are less that $\sqrt{0.8} c$ from Fig.~\ref{fig.6}.

\begin{figure}[htb]
	\centering
	\hspace*{-15pt}
	\includegraphics[scale=0.6]{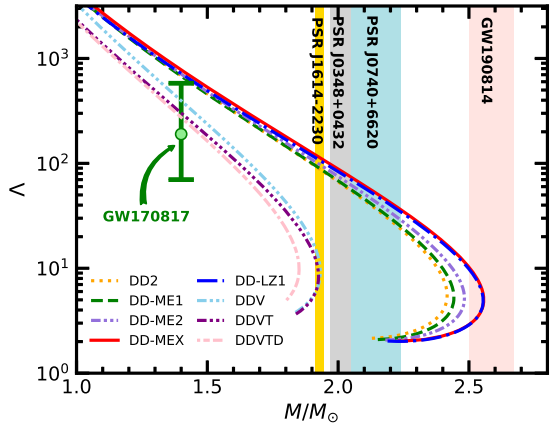}
	\caption{The tidal deformabilities from various DDRMF models as functions of neutron star mass. The mass regions of massive neutron stars are also plotted. }\label{fig.8}
\end{figure}

With the rapid developments of gravitational wave detectors, the tidal deformability of neutron star can be extracted from the BNS merger. It can be calculated with the Love number by solving a first-order differential equation, Eq.~\eqref{lneq}. In Fig.~\ref{fig.8}, the dimensionless tidal deformabilities, $\Lambda$, of neutron star as function of their masses from DDRMF models are shown. These dimensionless tidal deformabilities decrease with the neutron star mass and become very small at the maximum masses. Their values in softer group sets are significantly lower than those from other parameterizations, since $\Lambda\propto R^5$ approximately from Eq. \eqref{dt}. The radii of neutron star from the softer group EOSs are smaller comparing to the stiffer EOSs. The corresponding $\Lambda$ at $1.4 ~M_\odot$, $\Lambda_{1.4}$ are from $274.91$ to $390.01$, while recent analysis by LVC  gives  $\Lambda_{1.4}=190^{+390}_{-120}$ from GW170817~\citep{abbott2018}. Due to the larger radii and speeds of sound of neutron star in stiffer group EOSs, their $\Lambda$ are relatively higher and  $\Lambda_{1.4}$ are between $639.03$ and  $790.01$.  Furthermore, the tidal deformabilities at $2.0 ~M_\odot$ from these two types of EOSs have obviously differences. For the softer EOSs, $\Lambda$ almost approach to zero, while they are around $100$ for the stiffer EOSs at $2.0 ~M_\odot$. Once the BNS merger, whose components are around $2~M_\odot$, is more precisely measured by the advanced gravitational wave detectors in the future, the EOSs of neutron star can hopefully be determined well. Due to the large uncertainties in the present estimations of tidal deformability, we think that it cannot exclude the possibilities of stiffer EOSs with larger speeds of sound in neutron star, such as those from the stiffer group parameterizations. Therefore, the secondary compact object in GW190814 may be a neutron star.   

Finally, properties of neutron star, i. e., the maximum mass ($M_{\rm max}/M_{\odot}$), the corresponding radius ($R_{\rm max}$), the central density density($\rho_c$), the radius ($R_{1.4}$) and dimensionless tidal deformability ($\Lambda_{1.4}$) at $1.4 ~M_\odot$ from present DDRMF models are listed in Table~\ref{table.3}, respectively.
\begin{table}[htb]
	\footnotesize
	\centering
	\caption{Neutron star properties from various DDRMF models.}\label{table.3}
	\begin{tabular}{lrrrrrrrrrrr}
		\hline\hline
		&DD-LZ1    ~~&DD2       ~~&DD-ME1     ~~&DD-ME2     ~~&DD-MEX    ~~&DDV     ~~&DDVT       ~~&DDVTD  \\
		\hline
		$M_{\rm max}/M_{\odot}$         &2.5545    ~~&2.4168    ~~&2.4426     ~~&2.4829     ~~&2.5566    ~~&1.9317  ~~&1.9251     ~~&1.8507 \\
		$R_{\rm max}[\rm km]$           &12.178    ~~&11.826    ~~&11.885     ~~&12.012     ~~&12.274    ~~&10.336  ~~&10.023     ~~&9.850 \\
		$\rho_{\rm max}[\rm fm^{-3}]$   &0.786     ~~&0.845     ~~&0.832      ~~&0.813      ~~&0.777     ~~& 1.188  ~~&1.237      ~~&1.306 \\
		$R_{\rm 1.4}[\rm km]$           &12.864    ~~&12.938    ~~&12.931     ~~&12.961     ~~&13.118    ~~&12.195  ~~&11.511     ~~&11.396 \\
		$\Lambda_{\rm 1.4}$             &727.071   ~~&639.032   ~~&686.786    ~~&730.737    ~~&790.051   ~~&390.005 ~~&301.388    ~~&274.908\\
		\hline\hline
	\end{tabular}
\end{table}

%-------------------------------------------------------------------------------
	\section{Summaries and perspectives}\label{summary}
	   The latest density-dependent relativistic mean-field (DDRMF) parameterizations were systematically applied to investigate the properties of neutron star, i. e., DD2, DD-ME1, DD-ME2, DD-MEX, DD-LZ1, DDV, DDVT, and DDVTD sets. All of them were determined by fitting properties of spherical finite nuclei and have the same density-dependent function forms for meson coupling constants. Their densities, binding energies, incompressibilities, and symmetry energies at saturation points of symmetric nuclear matter are almost identical. 
	   
	   The equations of state (EOSs) of symmetric nuclear matter and pure neutron matter from present sets were separated into the softer type and stiffer one at high density region. The softer EOSs are generated by the DDV, DDVT, and DDVTD, whose coupling strengths of $\sigma$ and $\omega$ mesons are weaker comparing to other sets. Their vector and scalar potentials have comparable magnitudes, while the vector potentials are much larger than the scalar ones in stiffer EOSs given by DD2, DD-ME1, DD-ME2, DD-MEX, and DD-LZ1. Their pressures in symmetric nuclear matter at $2\sim4 \rho_{B0}$ were a little bit higher than the present constraints from heavy-ion collisions, while the softer EOSs satisfied these constraints.
	   
	   The TOV equation was solved using the EOSs of neutron star matter, where the nucleons and leptons are in conditions of beta equilibrium and charge neutrality, generated by present DDRMF models. The softer EOSs from DDV, DDVT, and DDVTD only can support the neutron stars with maximum masses around $1.90 ~M_\odot$ at  $10$ km and tidal deformabilities at $1.4~ M_\odot$, $\Lambda_{1.4}=274-390$.  The stiffer EOSs can generate very massive neutron stars around $2.5~M_\odot$. In particular, the DD-MEX and DD-LZ1 parameter sets even can produce neutron stars with masses of  $2.55~M_\odot$, which can explain the secondary object in GW190814 with a  mass of $2.50-2.67 ~M_\odot$. Furthermore, their radii at $1.4 ~M_\odot$  are also consistent with the constraints from NICER including the mass and radius simultaneous measurement, although their  $\Lambda_{1.4}$ were around $639-790$.
	   
	   In this investigation, we found that several parameterizations in DDRMF can provide very massive neutron stars due to the strong repulsive contributions from vector mesons at high density, which can describe ground state properties of finite nuclei exactly at the same time. The stiffer EOSs may slightly exceed the constraints of EOS from heavy-ion collisions and tidal deformability from GW170817. However, due to the strong model dependence of these two constraints and their large uncertainties, we can not exclude the possibility of the secondary object of GW190814 as a neutron star consisting of nucleons and leptons. We have shown that the stiffer EOSs give the dimensionless tidal deformability around $100$ for $2~M_\odot$ massive neutron star, while the softer ones less than $10$.  Therefore, the more precise measurement of dimensionless tidal deformability by the gravitational wave detectors will help us to determine the proper EOSs in the future.  The density dependence of coupling constant in DDRMF model provides a good mechanism, which can describe probably the finite nuclei and supranuclear matter concurrently.

\section{Acknowledgments}
This work was supported in part by the National Natural Science Foundation of China (Grants  No. 11775119, No. 11675083, and No. 11405116),  the Natural Science Foundation of Tianjin, and China Scholarship Council (Grant No. 201906205013 and No. 201906255002).

\end{document}